\documentclass[letterpaper,titlepage,12pt]{article}
\usepackage{setspace} % Always before hyperref -- Enable correct footnote link
\usepackage[colorlinks=true,urlcolor=blue,citecolor=magenta]{hyperref}
\usepackage{xcolor}
\usepackage{amssymb,amsmath,amsfonts}
\usepackage{epsfig}
\usepackage{epsfig}
\usepackage{graphicx}
\usepackage{epstopdf}
\usepackage[round]{natbib}
\usepackage{caption}
\usepackage{subcaption}
\usepackage{amscd}
\usepackage{amsthm}
\usepackage{csquotes}
\usepackage{latexsym}
\usepackage{float}
\usepackage{mathtools}
\usepackage[stable]{footmisc}
\usepackage{amsbsy}
\usepackage{bbm}
\usepackage[english]{babel}
\usepackage{psfrag}
\usepackage{tabularx}
\usepackage{csquotes}
\usepackage{colortbl}
\usepackage{braket}
\usepackage{subfloat}

\allowdisplaybreaks

\newcommand{\be}{\begin{eqnarray}}
\newcommand{\ee}{\end{eqnarray}}
\newcommand{\beq}{\begin{eqnarray}}
\newcommand{\eeq}{\end{eqnarray}}

\def\clock{{\count0=\time
 \divide\count0 60
 \ifnum\count0<10 0\fi\the\count0
 \multiply\count0 -60 \advance\count0 \time
 :\ifnum\count0<10 0\fi \the\count0
 }}
\newcommand{\timestamp}{{\small\vbox{\hbox{\tt\jobname.tex}
\hbox{\the\day/\the\month/\the\year, \clock}}}}

\begin{document}

\vskip 1.4 cm
\centerline{\LARGE \bf The Devil in the (Implicit) Details}
\vskip .25cm
\centerline{\large \bf On the AMPS Paradox and its Resolution}
\vskip 1.5cm
\centerline{\large {{\bf Enrico Cinti$^{1,2}$, Marco Sanchioni$^1$}}}
\vskip .8cm
\begin{center}
\sl $^1$ DISPeA, University of Urbino,\\
\sl Via Timoteo Viti 10, 61029 Urbino PU , Italy.\\
\end{center}
\begin{center}
\sl $^2$ Department of Philosophy, University of Geneva,\\
\sl 5, rue de Candolle, CH-1211 Gen\`eve 4, Switzerland.\\
\end{center}
\vskip 0.6cm
\centerline{\small\tt marco.sanchioni2@gmail.com, cinti.enrico@gmail.com}

\vskip .8cm \centerline{\bf Abstract} \vskip 0.2cm \noindent
The black hole information loss paradox has long been one of the most studied and fascinating aspects of black hole physics. In its latest incarnation, it takes the form of the firewall paradox. In this paper, we first give a conceptually oriented presentation of the paradox, based on the notion of causal structure. We then suggest a possible strategy for its resolutions and see that the core idea behind it is that there are connections that are non-local for semiclassical physics, which have to be taken into account when studying black holes. We see how to concretely implement this strategy in some physical models connected to the ER=EPR conjecture.


%%%%%%%%%%%%%%%%%%%%%%%%%%%%%%%%%%%%%%%%%%%%%%%%%%%%%%%%%%%%%%%%%%%%%%%%%%%%%%%%%%%%%%%%%%%%%%%%%%%%%%%%%%%%%%%%%%%%%%%%%%%%%%%%%%%%%%%%%%%%%%%%%%%%%%%%%%%%%%%%%%%%%%%%%%%%%%%%%%%%%%%%%%%%%%%%%%%%%%%%%%%%%%%%%%%%%%%%%%%%%%%%%%%%%%%%%%%%%%%%%%%%%%%%%%%%%%%%%%%%%%%%%%%%%%%%%%%%%%%%%%%%%%%%%%%%%%%%%%%%%%%%%%%%%%%%%%%%%%%

\tableofcontents

%%%%%%%%%%%%%%%%%%%%%%%%%%%%%%%%%%%%%%%%%%%%%%%%%%%%%%%%%%%%%%%%%%%%%%%%%%%%%%%%%%%%%%%%%%%%%%%%%%%%%%%%%%%%%%%%%%%%%%%%%%%%%%%%%%%%%%%%%%%%%%%%%%%%%%%%%%%%%%%%%%%%%%%%%%%%%%%%%%%%%%%%%%%%%%%%%%%%%%%%%%%%%%%%%%%%%%%%%%%%%%%%%%%%%%%%%%%%%%%%%%%%%%%%%%%%%%%%%%%%%%%%%%%%%%%%%%%%%%%%%%%%%%%%%%%%%%%%%%%%%%%%%%%%%%%%%%%%%%%

\section{Introduction}\label{1}
Throughout the years, black holes have proven to be one of the primary sources for clues towards a better understanding of the principles underlying a putative theory of Quantum Gravity (QG). In particular, Hawking's {calculation} of black hole radiation \cite{PhysRevD.14.2460}, and the subsequent development of the debate around the Black Hole Information Paradox (BHIP) {and the Page Time Paradox (PTP)} have played a pivotal role in clarifying their far-reaching impact on our understanding of physics. While black holes have been the ideal playground for studying many technical problems regarding quantum field theory in curved spacetime and semiclassical gravity, they play an even more critical role as catalysts for a better conceptual understanding of QG. For example, Hawking believed that BHIP showed that QG must be non-unitary \cite{PhysRevD.14.2460}, while AMPS believe that it shows that the equivalence principle is violated at the horizon \cite{Almheiri_2013}. \\
Philosophers did not ignore the importance of the conceptual study of black holes. Works of this type are \cite{Wallace:2017wzs} and \cite{10.2307/40072220}. However, both articles deal with variations of BHIP {and PTP}, which revolve around Hawking's original idea that black hole physics might be non-unitary (and ways to avoid this conclusion). {Nevertheless, most contemporary high-energy physicists are not usually concerned with the unitarity of black hole physics (which especially among string theorists is taken to follow from the AdS/CFT duality \cite{Maldacena:1997re,10.5555/2834415}, where unitarity is a standard feature of the boundary CFT), but rather with the structure of the interior of the black hole. The \textit{firewall paradox}, also known as the \textit{AMPS paradox} from the initials of its authors {\cite{Almheiri_2013}}, plays a central role.} Our goal {as philosophers of physics} is to study the conceptual foundations of the firewall paradox and to explore how {dropping an implicit assumption on the structure of spacetime, what we call \textbf{spacetime distinctness},} resolves it. {In particular, we highlight, by looking at concrete physical examples, how recent discussions regarding the resolution of the firewall paradox \cite{Papadodimas:2012aq,Maldacena:2013xja,Papadodimas:2015jra,Hayden:2018khn,Almheiri:2019psf,penington2019entanglement,Almheiri:2019hni,Almheiri:2019qdq} appear to rely crucially on this strategy.} We do not claim, of course, that our discussion is in any way exhaustive of the possible solutions to the firewall paradox.\footnote{For two examples of possible alternative resolutions see \cite{Harlow_2013,Rovelli_2019}.} We are also bracketing, for this paper, the various philosophical issues related to the proper definition and understanding of black holes \cite{Curiel:2018cbt}. \\
To clarify the AMPS paradox's conceptual structure and its resolution, we rely on the notion of \textit{causal structures} which we develop in \S\ref{2}. The upshot of our analysis is that the firewall paradox crucially depends on the assumption that relativistic locality, i.e. that only causal curves can carry causal influences, broadly understood in term of counterfactually robust correlations, is preserved in QG and that a natural way to resolve the paradox is to drop this assumption. \\
This work is of interest to both physicists and philosophers working on the foundations of black holes and QG. The conceptual insights provided by our analysis both serve to elucidate the philosophical foundations of the topic and help researchers in this field gain a better appreciation of the tools they are using. This way, we bridge the gap between the two communities, furthering the study of black holes as both a technical and conceptual topic.\\
The paper is structured as follows: in $\S\ref{2}$, we introduce the basic notion that we use throughout the rest of the paper to extract the firewall paradox's conceptual content: causal structures. In $\S\ref{3}$, we give a conceptually oriented introduction to the firewall paradox and study it in terms of causal structures. In $\S\ref{4}$, we study some concrete physical models connected to the ER=EPR conjecture and at the core of recent discussions of the firewall paradox, and clarify that they solve the paradox precisely by dropping \textbf{spacetime distinctness}. $\S\ref{5}$ then concludes.

%%%%%%%%%%%%%%%%%%%%%%%%%%%%%%%%%%%%%%%%%%%%%%%%%%%%%%%%%%%%%%%%%%%%%%%%%%%%%%%%%%%%%%%%%%%%%%%%%%%%%%%%%%%%%%%%%%%%%%%%%%%%%%%%%%%%%%%%%%%%%%%%%%%%%%%%%%%%%%%%%%%%%%%%%%%%%%%%%%%%%%%%%%%%%%%%%%%%%%%%%%%%%%%%%%%%%%%%%%%%%%%%%%%%%%%%%%%%%%%%%%%%%%%%%%%%%%%%%%%%%%%%%%%%%%%%%%%%%%%%%%%%%%%%%%%%%%%%%%%%%%%%%%%%%%%%%%%%%%%

\section{Causal Structures}\label{2}
We begin by introducing the notion that we employ throughout this article to analyze the firewall paradox:
\begin{itemize}\label{causal1}
\item[\textbf{(CS)}] \textbf{Causal Structure}: given a theory $T$, we say that the causal structure {according to} the theory $T$ is given by a set of spacetime regions/objects (with their physical state) and a relation $R$ which determines if two objects/regions of spacetime can or cannot be causally related.
\end{itemize}
\noindent To start, let us observe that when we talk about causal connections, this is done mostly for ease of exposition. By causality, we only mean that there are robust counterfactual connections between entities. No more robust notion of causality is assumed. The reader who prefers a stronger notion of causality is free to substitute for our talk of causality talk of robust counterfactual connections.\footnote{{Note however that, for our discussion, one should use \textit{robust counterfactual connections} when defining causal structures; otherwise entanglement would not fall under this notion.}}\\
To get a feel for this notion, let us use it in the case of General Relativity (GR). Here, the theory $T$ is just GR, and the objects of the theory are spacetime points.\footnote{{Note that it is not apparent how to interpret GR. For ease of exposition, we speak explicitly of spacetime points. Nonetheless, it should be possible to carry over this discussion for more refined approaches to GR's ontology.}} To capture the causal structure of the theory, then, we can define a relation $R_{LC}$ of \textit{being connectable by a causal curve}. This relation obtains between two spacetime points $p$ and $q$ if and only if a causal curve can connect them. Observe that this relation captures GR's locality properties since it entails that only spacetime points connectable by a causal curve can be in causal contact.\\
Moreover, entanglement relations, since they define robust counterfactual connections, as observed in \cite{Maudlin2002-MAUQNA}, also define a causal structure. In this case, the background theory would be QM or QFT, and the objects would be quantum systems. The relation defining the causal structure would then be $R_E$, i.e.\ \textit{being entangled with}, obtaining between two quantum systems $A$ and $B$ if and only if they are entangled.

%%%%%%%%%%%%%%%%%%%%%%%%%%%%%%%%%%%%%%%%%%%%%%%%%%%%%%%%%%%%%%%%%%%%%%%%%%%%%%%%%%%%%%%%%%%%%%%%%%%%%%%%%%%%%%%%%%%%%%%%%%%%%%%%%%%%%%%%%%%%%%%%%%%%%%%%%%%%%%%%%%%%%%%%%%%%%%%%%%%%%%%%%%%%%%%%%%%%%%%%%%%%%%%%%%%%%%%%%%%%%%%%%%%%%%%%%%%%%%%%%%%%%%%%%%%%%%%%%%%%%%%%%%%%%%%%%%%%%%%%%%%%%%%%%%%%%%%%%%%%%%%%%%%%%%%%%%%%%%%

\section{Black Hole Paradoxes\footnote{For background material on black hole physics see \cite{Harlow:2014yka}.}}\label{3}
{In this section, we start by briefly reviewing various paradoxes regarding black holes to provide background for our subsequent discussion of firewalls. In particular, we follow \cite{Wallace:2017wzs} in distinguishing between two main paradoxes: BHIP (\S\ref{3.1}) and PTP (\S\ref{3.2}). Then we introduce the firewall paradox (\S\ref{3.3}) and discuss its interpretation in terms of causal structures (\S\ref{3.4}).}

\subsection{Black Hole Information Paradox}\label{3.1}
\begin{figure}
\centering
\includegraphics[scale=0.6]{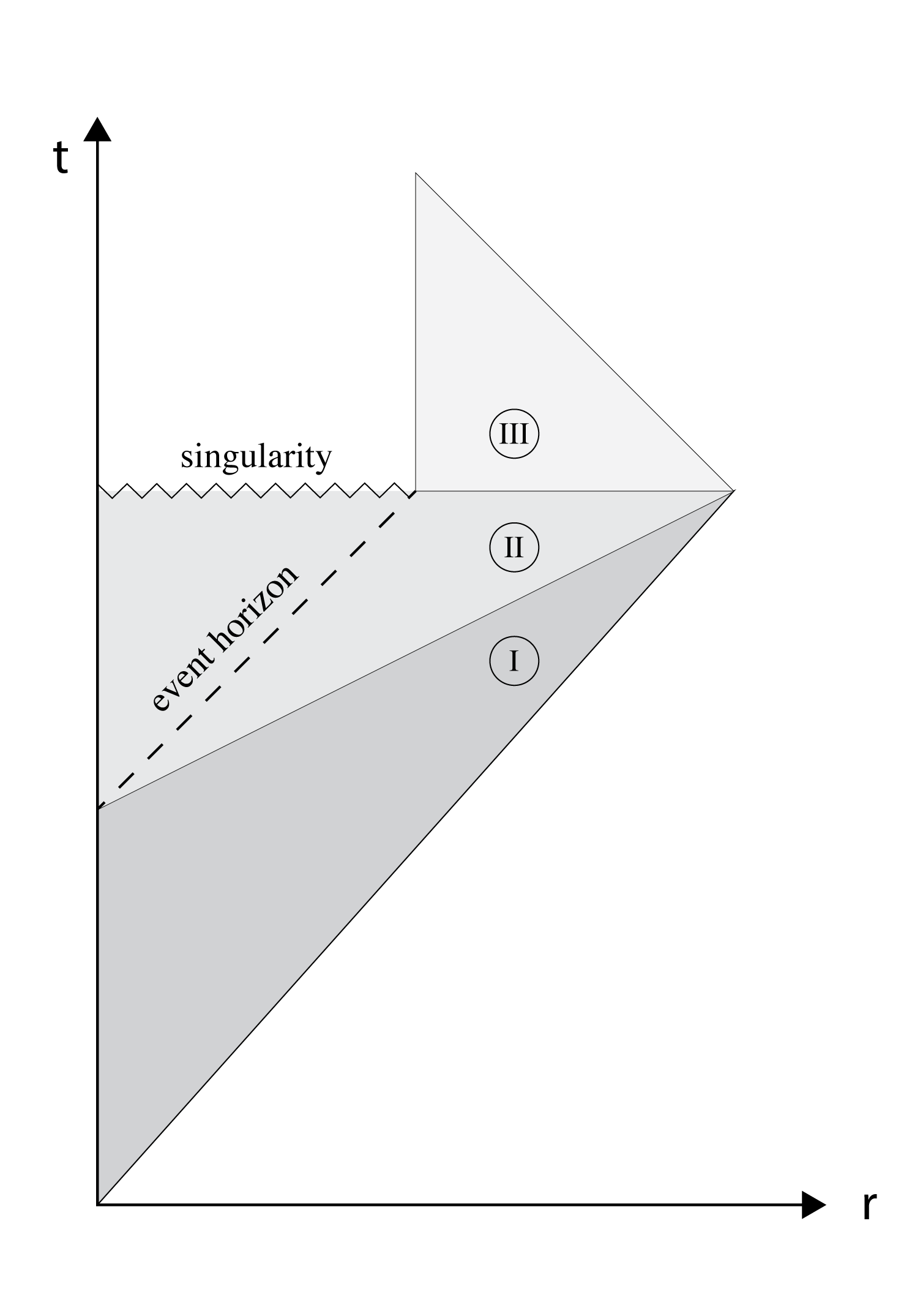} 
\caption{Spacetime diagram of an evaporating black hole.}
\label{fig:BHIP_Penrose}
\end{figure}

BHIP concerns the non-unitarity of the complete evaporation of a black hole.\footnote{See \cite{10.2307/40072220}.} Consider an evaporating black hole as in {Figure} \ref{fig:BHIP_Penrose}. {In the coordinate frame of an asymptotic observer,\footnote{{In this paper, unless specified, all statements are made in such a coordinate frame.}}} we can distinguish three regions in the diagram: (I) the pre-formation region, i.e. the collapse of a star into a black hole; (II) the evaporation region, i.e. the region in which the black hole starts evaporating due to the emission of Hawking quanta; (III) the post-evaporation region, i.e. the region in which the black hole has evaporated. 
Hawking showed that the quantum state outside the black hole in the region (II) is in a mixed state, i.e. that the state is perfectly thermal \cite{PhysRevD.14.2460}. In other words, the quantum state of the Hawking radiation is determined only by macroscopic properties of the evaporating black hole, i.e. its mass, charge and angular momentum. In turn, Hawking radiation being exactly thermal also means that {we cannot retrodict the state of the star collapsing into the black hole. The information encoded in this state should then be stored elsewhere, and the only possibility is the interior of the black hole.\footnote{Understanding the meaning of information in BHIP is certainly important. However, we bracket this issue since nothing of what we are going to say hinges on it.} Nevertheless, since the black hole eventually completely evaporates, this is only possible in region (II). Indeed, there are no slices in region (III) from which we can retrodict the state of the collapsing star, and thus there cannot be unitary evolution from the region (I)$+$(II) to region (III). It is a controversial issue whether one should take BHIP as an argument for the non-unitarity of black hole evaporation \cite{Maudlin:2017lye,Unruh:1995gn,Unruh:2017uaw} or as a true paradox. However, we do not take a specific stance on this issue since nothing in this article depends upon it.}

\subsection{Page Time Paradox}\label{3.2}
\begin{figure}[h!]
\centering
\includegraphics[scale=0.8]{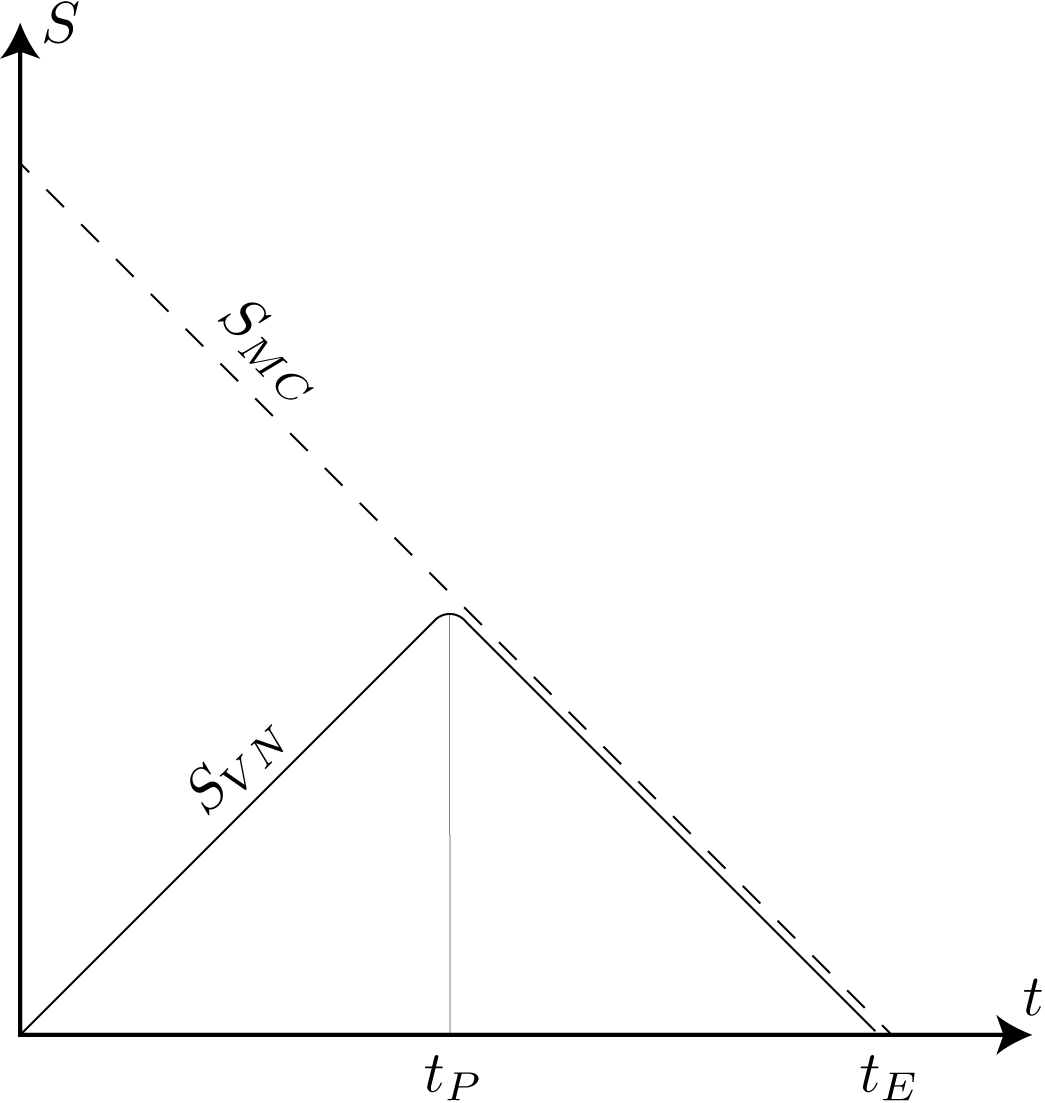} 
\caption{The Page curve for an evaporating black hole.}
\label{fig:PageCurve}
\end{figure}

Page presents a different paradox which applies long before the evaporation time {\cite{Page_1993}.\footnote{See \cite{Wallace:2017wzs} for a detailed philosophical discussion.}} Consider the black hole to be a thermodynamic system in a pure state, which we describe via a microcanonical ensemble, i.e. an ensemble whose (thermodynamic) entropy $S_{MC}${, where $MC$ stands for microcanonical,} is given by:
\beq\label{eq:microS}
S_{MC}\left(E(t)\right) = \log \dim \mathcal H\left[E(t)\right],
\eeq
where $\mathcal H\left[E(t)\right]$ is the Hilbert space of the system at some time $t$ and energy $E(t)$.\footnote{{For a review of quantum statistical mechanics see \cite{mussardo2010statistical}.}} Moreover, given a composite quantum system, define the von Neumann entropy\footnote{Also known as \textit{fine-grained} entropy or \textit{entanglement} entropy. We use these terms interchangeably in what follows.} of one of its subsystems as
\beq\label{eq:microV}
S_{VN}=-\mbox{Tr}\left( \rho\log\rho\right)
\eeq
where $\rho$ is the reduced density matrix of that subsystem.\\
Since the black hole was in a pure state before evaporation, then by unitarity at any time $t$ after the evaporation started the composite system of the black hole and the Hawking radiation should be in a pure state. If one thinks of the radiation and the black hole as subsystems, then their Von Neumann entropies give the amount of entanglement between them and must be the same. We thus simply speak of $S_{VN}$. In particular, $S_{VN}$ increases with the emission of Hawking quanta entangled with the interior. Moreover, we assume that each Hawking quantum is entangled with an interior mode, to maintain the black hole plus radiation system in a pure state. Therefore, if black hole evaporation is unitary, $S_{MC}$ bounds $S_{VN}$, since $S_{MC}$ is proportional to the dimension of the Hilbert space of the black hole \eqref{eq:microS}, and there cannot be more interior modes that the dimensionality of the black hole's Hilbert space. We call this bound the \textit{Page bound}:
\beq\label{eq:bound}
S_{VN} \leq S_{MC}.
\eeq
The black hole cools down through evaporation and loses energy emitting Hawking quanta in a perfectly mixed state, which means that the microcanonical entropy $S_{MC}$ decreases over time. On the other hand, $S_{VN}$ increases with the number of emitted Hawking quanta, since each Hawking quanta is entangled with an interior mode.\footnote{Moreover, also the microcanonical entropy of the Hawking radiation increases with time since its microcanonical entropy is proportional to the number of Hawking quanta.} Therefore there must be a time $t_P$, called the Page time, at which the bound \eqref{eq:bound} saturates, which means that all interior modes are entangled with a Hawking quantum. The Page time $t_P$ is also when the microcanonical entropy of the radiation becomes bigger than the microcanonical entropy of the black hole. The resulting curve, which initially grows with $S_{VN}$ and then decreases with $S_{MC}$, is the \textit{Page curve} (see {Figure} \ref{fig:PageCurve}).
 \\
Consequently, since after $t_P$ there cannot be enough interior modes to keep the composite system of black hole and radiation in a pure state, one would expect the violation of the bound \eqref{eq:bound} and the non-unitarity of the evaporation process. The way to avoid this conclusion is for the late-time Hawking radiation\footnote{In what follows the words \textit{late} and \textit{early} refer respectively to radiation emitted \textit{after} and \textit{before} the Page time $t_P$.} not to be entangled with an interior mode, but with something else. The only possibility is that it is entangled with the \textit{early-time radiation}, i.e. the Hawking radiation emitted at times $t<t_P$. In this way, the early radiation purifies the late, keeping the state of the black hole plus radiation system pure. Moreover, the entanglement entropy of the Hawking radiation decreases after $t_P$, respecting the bound \eqref{eq:bound}. {The entanglement between early and late radiation implies that the Hawking radiation is not perfectly thermal, since there are non-trivial correlations among its constituents. Nevertheless, Hawking showed that the radiation \textit{is} perfectly thermal.} Therefore we have a contradiction. This is the nature of PTP: the inconsistency between the prediction of \textit{naive}\footnote{Here by \textit{naive} we simply mean the picture of quantum fields living on the smooth Lorentzian manifold of GR, as used in \cite{PhysRevD.14.2460}.} semiclassical gravity (Hawking's calculation) and black hole statistical mechanics (the entropy bound \eqref{eq:bound}). PTP occurs long before the evaporation time. Indeed, it occurs at the Page time $t_p$, approximately half of the evaporation time for a Schwarzschild black hole. \\
{A resolution of PTP is offered by the AdS/CFT correspondence, within which one can show that the prediction of black hole statistical mechanics is the correct one.\footnote{For a defence of this conclusion see \cite{Wallace:2017wzs}. In particular, assuming that the correspondence is valid, black hole physics can be shown to be unitary, implying that Hawking radiation cannot be thermal.} Thus black hole entropy follows the Page curve, grows with time until $t_P$ and decreases afterwards, and the bound \eqref{eq:bound} is not violated.\footnote{As Susskind said \enquote{South America wins the war}\cite{susskind2008black}, referring to the nationality of Juan Maldacena, the author of \cite{Maldacena:1997re}, the first article on AdS/CFT.}}

\subsection{The Firewall Paradox}\label{3.3}
{While the discussion around PTP has centred around arguments in favour or against the unitarity of black hole physics, said (non) unitarity is not exhaustive of the range of possible black hole paradoxes. In particular, in this section, we describe the firewall paradox developed in \cite{Almheiri_2013} (henceforth we refer to the authors of this article as AMPS), which threatens the possibility of constructing a consistent theory of the interior. To start, let us follow AMPS in taking the following four postulates (originally formulated in \cite{Susskind_1993} and widely accepted among high-energy physicists\footnote{{At least among physicists in the string theory community. Note that these postulates were originally defined in the context of black hole complementarity. However, they are not strictly tied to this specific proposal. {Rather, they are taken to be definitive of what a sensible theory of black hole physics should look like.}}}) to be reasonable assumptions that any theory of black holes should satisfy:}
\begin{itemize}
\item[Postulate 1] The process of formation and evaporation of a black hole, as viewed by a distant observer, can be described by a unitary S-matrix encoding the evolution from infalling matter to outgoing Hawking-like radiation.
\item[Postulate 2] Outside the stretched horizon\footnote{An horizon at a distance of a Planck length from the true event horizon. It is time-like.} of a massive black hole, physics can be described to a good approximation by a set of semiclassical field equations.
\item[Postulate 3] To a distant observer, a black hole appears to be a quantum system with discrete energy levels. The dimension of the subspace of states describing a black hole of mass $M$ is the exponential of the Bekenstein entropy $S_{MC}$.
\item[Postulate 4] A freely falling observer experiences nothing out of the ordinary when crossing the horizon until the singularity is approached. Another way to say this is that no observer ever detects a violation of the known laws of physics.
\end{itemize}
AMPS showed that these four postulates are inconsistent. Indeed, take an evaporating black hole and consider its Hawking radiation. Divide the black hole plus radiation system into three subsystems: early radiation $E$, late radiation $L$ and the interior partners to the late radiation $B$. As we have seen in the previous section, the fact that the Hawking radiation is in a pure state means that the late radiation $L$ and the early radiation $E$ should be entangled. Indeed, a subsystem of a bipartite system in a pure state is maximally entangled with its counterpart. Therefore, since the combined system of early and late radiation is pure (a consequence of Postulate 1), then the early radiation and the late radiation should be maximally entangled. Moreover, if there is no drama at the horizon (Postulate 4), the late radiation $L$ is fully entangled with the modes behind the horizon $B$. Indeed, for an infalling observer, the geometry near the horizon can be locally identified with a Rindler horizon via a coordinate transformation:
\beq
\tau=\frac{t}{4M}\quad,\quad \rho =2\sqrt{2m(r-2M)}\quad,\quad \tilde x =\left(2M\theta,2M\phi\right),
\eeq
which turns the Schwarzschild metric into 
\beq
ds^2 \approx -\rho^2 d\tau^2 +d\rho^2+d\tilde x^2,
\eeq
the metric corresponding to Rindler spacetime.\footnote{{Note that, here, $d\tilde x^2 = d\tilde x_1^2+ d\tilde x_2^2$, where $x_1 = 2M\theta$ and $x_2 = 2M\phi$.}} Fields in the left Rindler wedge are maximally entangled with fields in the right Rindler wedge. Therefore a mode $L$ outside the horizon must be maximally entangled with a mode $B$ inside the horizon. Thus the composite system, made of $E$, $L$ and $B$, has two features: $L$ is maximally entangled with both $E$ and $B$. 
\begin{figure}
\centering
\includegraphics[scale=0.8]{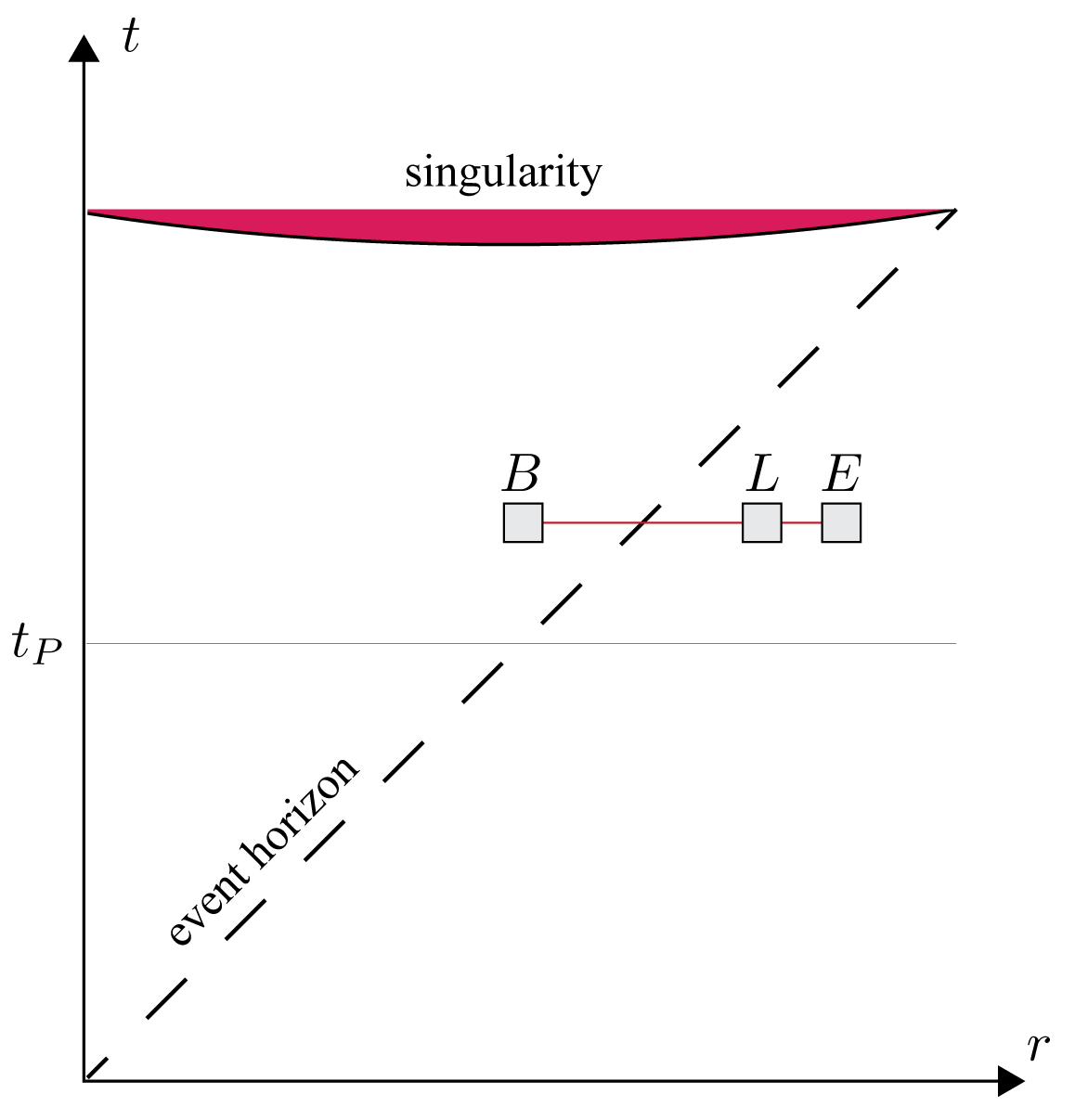} 
\caption{A sketch of the basic scenario behind the firewall paradox. $B$ is an interior mode, $L$ is a mode of late Hawking radiation, while $E$ is a mode of early Hawking radiation. The red lines represent relations of maximal entanglement.}
\label{fig:AMPS}
\end{figure} \\
However, a fundamental property of entangled systems is what in quantum information is called the monogamy of entanglement: a quantum system can be maximally entangled with \textit{only one} quantum system at a time. {Equivalently, monogamy of entanglement says that there is an upper bound on the independent degrees of freedom with which a given quantum system can be entangled, given by the number of independent degrees of freedom of the quantum system itself.} Thus the four postulates are in contradiction with the monogamy of entanglement. This contradiction is known as the firewall paradox.\footnote{An equivalent formulation of the paradox is as a violation of strong subadditivity of entanglement entropy, i.e.\ $S_{AB}+S_{BC}\geq S_B + S_{ABC}$.} AMPS proposed that $L$ and $B$ are not entangled, but are instead in a product state. Since $L$ and $B$ being in a product state means that close to the horizon Rindler space is not a good model for spacetime, the assumption that an observer does not see anything out of the ordinary there [Postulate 4] is not reliable. Indeed, if there is no entanglement between $L$ and $B$, $B$ is not entangled with anything. On the other hand, $L$ is still entangled with $E$, which purifies it. Therefore the state of $B$ is characterized by its reduced density matrix, which is thermal.\footnote{The thermal nature of the state of $B$ is because, since we want the entangled and non-entangled descriptions to be locally indistinguishable, to describe the state of $B$ we use the reduced density matrix of the maximally entangled state of the composite system of $B$ and $L$. It is a well-known fact that the reduced density matrix of a subsystem of a maximally entangled system is completely thermal.} The thermal nature of $B$ can be suggestively described as a wall of fire, hence the notion of the \textit{firewall}. \\
The existence of firewalls implies that a non-trivial theory of the interior is impossible since the interior would always be characterized by the same thermal state, i.e.\ the firewall. As such, no interesting theory of the black hole interior would be possible in the presence of firewalls.

\subsection{AMPS and Causal Structures}\label{3.4}
Let us now clarify where the tension lies in the firewall argument of AMPS. The standard formulation of the paradox is useful in showing the incompatibility of the four postulates. What remains unclear is where precisely the conflict giving rise to the incompatibility between relativity and quantum theory lies. To make clear what is the conceptual origin of this conflict, we make use of the notion of causal structure. \\
Beyond the four postulates listed, the fact that $B$ and $E$ are separate systems plays a fundamental role in the paradox. For black holes, the distinctness between $B$ and $E$ is justified because $B$ lies behind the horizon of the black hole. $B$ is thus causally isolated from $L$ and $E$ since the exterior and interior of the black hole are spacelike separated. However, this is not an assumption as innocent as it might sound: it is equivalent to the claim that relativistic notions of separability and locality, based on spacelike and causal connections, are retained in the regime of QG. Let us focus on this point, as it will be crucial in the rest of this paper. \\
As we have mentioned, to ensure the violation of monogamy, and thus the firewall paradox, one has to regard $B$ and $E$ as distinct systems, {where by distinct we mean that their degrees of freedom are independent. This fact follows from monogamy of entanglement giving an upper bound on the number of independent degrees of freedom with which a certain quantum system, in our case $L$, can be entangled.} Indeed, (i) we know that $B$ and $E$ are spacelike related since one is in the interior and the other in the exterior of the black hole. Furthermore, (ii) from the study of QFT (and in particular of Algebraic QFT\footnote{For reviews of AQFT see \cite{halvorson2007algebraic,haag2012local}.}) we have come to accept that if two algebras of observables\footnote{Note that its algebra of observables identifies a system.} are mutually commutative, then they represent two distinct systems,\footnote{Though see \cite{Earman2014-EARRCI} for some subtleties about this claim.} {since, when this is the case, their degrees of freedom are completely independent.} From the axiom of \textit{microcausality}, (iii) we have that the algebras of observables connected to two spacelike related regions must commute, i.e. $\left[\mathfrak{N}(\mathcal X),\mathfrak{N}(\mathcal X')\right]=0$, where $\mathfrak{N}(\mathcal X)$ and $\mathfrak{N}(\mathcal X')$ are the algebras of observables associated to two spacelike related regions $\mathcal X$ and $\mathcal X'$. Thus, from (i)-(ii)-(iii), the two systems must be distinct.\\
Note however that the microcausality axiom encodes the locality properties of classical relativistic spacetime since it relies explicitly on the notion of spacelike separation, which we have no guarantee will be retained at the level of quantum spacetime. Thus we have to assume that this notion of locality, developed for relativistic spacetimes, can be extended seamlessly beyond GR. If this were not the case, then $B$ and $E$ might not be distinct at the quantum level, opening the door to resolving the paradox by observing that there is no violation of monogamy since $B$ and $E$ are not distinct systems {which implies that their degrees of freedom are not independent and that $L$ is entangled with fewer degrees of freedom than those manifest in the semiclassical description} (more on this in $\S\ref{4}$).\\
We have thus seen that to the four postulates above, AMPS (implicitly) add a fifth one:
\begin{itemize}\label{distinctness}
\item[\textbf{(SD)}] \textbf{Spacetime Distinctness:} spacelike separated systems are distinct, i.e. mutually commuting.
\end{itemize}
\noindent
With this in mind, we can then reformulate the paradoxical conclusion of AMPS as follows: in a black hole spacetime, we have three subsystems $L$, $B$ and $E$, such that $B$ and $L$ are distinct systems, i.e. $\mathfrak{N}(B)$ commutes with $\mathfrak{N}(L)$, and they violate the monogamy principle, i.e. $L$ is maximally entangled with both $B$ and $E$.\\
To better keep track of the various moving parts of the AMPS argument, and to elucidate its conceptual content, let us recall the definition of causal structure:
\begin{itemize}\label{causal2}
\item[\textbf{(CS)}] \textbf{Causal Structure}: given a theory $T$, we say that the causal structure according to the theory $T$ is given by a set of spacetime regions/objects (with their physical state) and a relation $R$ which determines if two objects/regions of spacetime can or cannot be causally related.
\end{itemize}
{To recast the AMPS paradox in terms of causal structures, let us take as our objects quantum systems and let us consider two different relations, which define two different causal structures.} One is the relation $R_{LC}$ of \textit{being connectable by a causal curve} that we have encountered in $\S\ref{2}$, which defines what we call the causal structure of spacetime. The other is the relation $R_{ME}$ of \textit{being maximally entangled}, which defines what we call the causal structure of entanglement, and is a slight restriction of the relation $R_E$ seen in $\S\ref{2}$. The core claim of AMPS is then that, under the four postulates detailed in $\S\ref{3.3}$, there are three quantum systems $L$, $B$ and $E$ such that $R_{ME}(L,B)$ and $R_{ME}(L,E)$, and $B$ and $E$ are distinct, which follows from $\neg R_{LC}(B,E)$, i.e. $B$ and $E$ are spacelike related. Our four assumptions (supplemented with \textbf{spacetime distinctness}) are then in violation of the monogamy of entanglement (see {Figure} \ref{fig:AMPS}). In other words, combining the causal structure of spacetime and entanglement by simply superimposing them leads to a contradiction. To resolve this paradox, one has three possibilities:
\begin{itemize}
\item[(i)] accept $\neg R_{ME}(L,B)$, i.e.\ $L$ and $B$ are not entangled. This is AMPS' answer and implies a firewall at the horizon. 
\item[(ii)] accept $\neg R_{ME}(L,E)$, i.e.\ $L$ and $E$ are not entangled. This answer is equivalent to Hawking's calculation and implies the non-unitarity of black hole evaporation.
\item[(iii)] accept that $B$ and $E$ are not distinct. This answer implies, as we see in $\S\ref{4}$, the modification of the causal structure of spacetime.
\end{itemize}
The solution that we study in this paper is (iii). Let us briefly remark why we think it is more promising than (i) and (ii). First of all, (i) entails the violation of the equivalence principle of GR since it implies that freely falling observers do not experience the gravitational vacuum at the horizon but instead meet the firewall. Since the equivalence principle is one of the fundamental insights of GR, one should be careful about renouncing it. As regards (ii), there is a violation of unitarity. However, the non-unitarity in Hawking's calculation comes from Plank scale effects which take place at the end of the evaporation process. On the other hand, the AMPS paradox occurs long before this time, when semiclassical gravity and effective field theory are still approximately valid descriptions. As such, it is unclear how this violation of unitarity might arise in this context. Moreover, as we have observed at the end of \S\ref{3.2}, unitarity seems to be a core aspect of quantum theory and to be retained in quantum gravity, at least insofar as one can trust arguments based on the AdS/CFT correspondence. As such, both (i) and (ii) seem to imply the violation of well justified, though not immune from revision, principles. On the other hand, (iii) would imply, as we remarked above, that the relativistic notion of locality does not apply to quantum spacetime.\footnote{{Note that, while they both imply some violation of relativistic locality, this approach differs significantly from that of Giddings \cite{Giddings:2012gc,Giddings:2013kcj}. Indeed, Giddings's approach implies the existence of non-local interactions at the level of semiclassical, effective field theory, something which is not implied by (iii) which only requires that at the quantum gravitational level, but not necessarily at the semiclassical level, $B$ and $E$ cannot be treated as distinct systems. This difference is particularly clear in the language of causal structures since Giddings's proposal would imply the acceptance of $R_{LC}(B,E)$, i.e.\ $B$ and $E$ being causally connectable, which is not necessary for (iii). Indeed, in order to ensure the violation of \textbf{(SD)} it is crucial that $B$ and $E$ are space-like related, and thus $\neg R_{LC}(B,E)$. As a matter of fact, as observed in \cite{Harlow:2014yka}, Giddings's proposal is better characterized as a type of firewall, though from our classification it appears to be somewhat different from the firewall proposed in AMPS. Moreover, Giddings's proposal seems to imply either the violation of the laws of black hole thermodynamics \cite{Almheiri:2013hfa} or modifications of the Schwarzschild geometry away from the horizon \cite{Giddings:2013vda,Giddings:2014ova}, both consequences which are not shared by (iii).}} However, the justification for thinking this seems to be shakier than the other principles discussed. To our knowledge, there is no explicit argument in defence of this claim. As such, we take it that before accepting (i) or (ii), one should at least test the viability of (iii), as it requires milder revisions to the fundamental principles of physics than its alternatives.\\
Observe that the causal structure formulation allows us to clarify the dependence of the paradox on a precise notion of when two systems are distinct, something that is not evident from the formulation of AMPS. Furthermore, this way of expressing the paradox allows us to connect explicitly with previous discussions on the supposed non-local character of entanglement \cite{jarrett1984physical,Maudlin2002-MAUQNA}. {This feature} is traditionally associated with having spacelike separated systems which are however entangled. In the language of causal structures, this means that for two quantum systems $A$ and $B$, it can be the case that $R_{ME}(A,B)$ while $\neg R_{LC}(A,B)$. While this connection has been the source of many conceptual puzzles, results such as the no signalling theorem and discussions such as that of \cite{jarrett1984physical} have convinced philosophers and physicists alike that entanglement cannot violate locality in any observable way. Since entanglement does not violate locality, one might naively expect that the two causal structures can be combined by superimposing them without modifying one or the other. What AMPS show is that this expectation cannot survive in the context of quantum black holes. While the single entanglement connections between pairs of spacelike separated quantum systems are not problematic, the overall pattern of spacetime and entanglement connections remains incompatible. Indeed, one can generate situations, such as that of AMPS, where there are three or more distinct\footnote{Which, as we have observed at the beginning of this section, is justified in terms of their being spacelike separated.} maximally entangled quantum systems, something which should not be possible according to the rules of quantum entanglement.\\
Entanglement and spacetime structures are not compatible, despite the careful considerations that led some to think they were. We lacked the theoretical tools to capture the consequences of this incompatibility, theoretical tools that were finally developed in the study of quantum black holes, and put to use by AMPS. Understanding the extent of this incompatibility, and showing how to overcome it, is the task of any resolution of the firewall paradox.

%%%%%%%%%%%%%%%%%%%%%%%%%%%%%%%%%%%%%%%%%%%%%%%%%%%%%%%%%%%%%%%%%%%%%%%%%%%%%%%%%%%%%%%%%%%%%%%%%%%%%%%%%%%%%%%%%%%%%%%%%%%%%%%%%%%%%%%%%%%%%%%%%%%%%%%%%%%%%%%%%%%%%%%%%%%%%%%%%%%%%%%%%%%%%%%%%%%%%%%%%%%%%%%%%%%%%%%%%%%%%%%%%%%%%%%%%%%%%%%%%%%%%%%%%%%%%%%%%%%%%%%%%%%%%%%%%%%%%%%%%%%%%%%%%%%%%%%%%%%%%%%%%%%%%%%%%%%%%%%

\section{No Firewall on the Horizon}\label{4}
{In this section, we analyze route (iii)'s resolution of the firewall paradox. Before delving deeper into this topic, however, it is useful to recall the basic setup of \S\ref{3}. First of all, we have begun in \S\ref{3.1} and \S\ref{3.2} by reviewing the two central paradoxes in black hole physics: BHIP and PTP. In particular, PTP shows the incompatibility of unitary evaporation, as encoded in the Page curve of the black hole entropy, with Hawking's original description of black hole radiation \cite{PhysRevD.14.2460}. We have also seen that AdS/CFT has been argued to show that unitary evaporation ultimately is the correct option. However, as shown in \cite{Almheiri_2013}, unitarity is still not sufficient to give a consistent description of black holes and, in particular, of the interior. Indeed, one runs into the \textit{firewall paradox} (\S\ref{3.3}): from four reasonable postulates, one can derive a situation violating the monogamy of entanglement. Furthermore, we have proposed to study the conceptual structure of the firewall paradox in terms of \textit{causal structures} (\S\ref{3.4}). In this framework, the paradox arises as an incompatibility between the causal structures of entanglement and spacetime. In particular, we have seen that a fundamental ingredient for the paradox to arise is \textbf{(SD)}, which is equivalent to the claim that the spacetime causal structure of GR is imported in the final theory of QG. \\
While AMPS have suggested that one should resolve the paradox by positing that the infalling observer encounters a firewall of high energy quanta at the horizon, retaining \textbf{(SD)} at the cost of the equivalence principle, such a way to resolve the paradox is not the only one possible. In particular, we have argued in \S\ref{3.4} that dropping \textbf{(SD)} is a viable approach to resolving the paradox. In this section, we are going to {introduce} some concrete physical models which {are at the centre of recent discussions among physicists regarding the resolution of the firewall paradox \cite{Papadodimas:2012aq,Maldacena:2013xja,Papadodimas:2015jra,Hayden:2018khn,Almheiri:2019psf,penington2019entanglement,Almheiri:2019hni,Almheiri:2019qdq} and argue that they} avoid the paradox by instantiating the violation of \textbf{(SD)}. {This analysis serves to show that these various approaches all share the same basic conceptual strategy and to better appreciate its implications}. These models have been constructed in the context of AdS/CFT\footnote{Note that, in what follows, we only use AdS/CFT as a mathematical tool to construct specific physical models. In particular, we do not discuss any of the philosophical issues connected to holographic dualities \cite{deHaro:2015pia}.} and, in particular, they are instances of the ER=EPR conjecture {\cite{Maldacena:2013xja}}.\footnote{{Here, ER stands for Einstein and Rosen, from the seminal article \cite{PhysRev.48.73} introducing wormholes, while EPR stands for Einstein, Podolsky, and Rosen from the article \cite{PhysRev.47.777} which first pointed out the EPR paradox and the non-local character of entanglement.}} However, let us stress that these models do not strictly depend on the conjecture's truth but rather provide evidence for it since they can be constructed (somewhat) independently of ER=EPR. Ultimately we will see that the non-local connections characteristic of the ER=EPR conjecture, and concretely instantiated in these models, engineer the violation of \textbf{(SD)} in which we are interested. Thus how ER=EPR avoids the firewall paradox is a paradigmatic instance of route (iii)'s strategy for resolving the paradox. Insofar as the reader is sanguine regarding the prospects of ER=EPR, our discussion provides evidence that the paradox's solution lies in the violation of \textbf{(SD)}. For the reader to be more sceptical about the prospects of these approaches, we show that the violation of \textbf{(SD)} resolves the paradox in certain specific models and clarify what this resolution implies while leaving open if the same strategy applies to more general cases.}\\
We proceed in \S\ref{4.1} by discussing the instructive, though not realistic, case of the eternal AdS black hole where we can see how ER=EPR undermines \textbf{(SD)} fundamental to the AMPS paradox. In \S\ref{4.3}, we then move to the case directly relevant to the AMPS paradox, that of an evaporating black hole formed from gravitational collapse.\footnote{Further constructions, which implement the same idea and apply to generic black holes in AdS are presented in \cite{Papadodimas:2015jra,almheiri2018holographic}. Note that these {constructions} crucially rely on the notion of \textit{state-dependence}, originally introduced in \cite{Papadodimas:2012aq}.}

\subsection{Eternal Black Holes}\label{4.1}
Let us start from the case of eternal\footnote{By eternal here we mean a black hole which has always existed and will always exist. Thus, the black hole has not formed via gravitational collapse, and also it is not subject to evaporation. In AdS, the non-evaporation of the black hole is due to the reflecting boundary conditions of the spacetime, which mean that the Hawking radiation emitted from the black hole bounces back inside the black hole upon reaching the boundary of AdS spacetime. Thus, the black hole is in equilibrium with exterior spacetime and does not evaporate. To make an AdS black hole evaporate, one needs to have absorbing boundary conditions, which we discuss in $\S\ref{4.3}$.}  AdS black holes, following the treatment of \cite{Maldacena:2013xja}. While these are not the evaporating black holes generated from gravitational collapse, for which the AMPS paradox applies, they can still be instrumental in testing and developing ideas regarding black holes' structure. In particular, the purpose of starting from the case of a two-sided black hole\footnote{By a two-sided black hole we mean a physical system with two event horizons. As such, a two-sided black hole has two exterior regions, usually called the \textit{left} and the \textit{right} exterior. Eternal AdS black holes are two-sided black holes.} is that it serves a useful pedagogical role since it allows to phrase various questions regarding the interior structure of black holes to a level of precision hard to attain in the context of standard, evaporating, one-sided black holes.\footnote{By a one-sided black hole we mean a physical system with one event horizon. As such, a one-sided black hole has only one exterior region.} It is thus useful to start from this most basic case. \\
Eternal AdS black holes are holographically dual to a couple of entangled CFTs (called the \textit{left} and \textit{right} CFT depending on which of the two exterior regions of the AdS black hole they describe) in the so-called \textit{thermofield double} state: 
\beq
\ket\psi = \sum_j e^{-\beta E_j/2}{\ket j}_L \otimes {\ket j}_R
\eeq
where $E_j$ are the energy levels of the CFT and $\beta=1/T$, where $T$ is the temperature of the black hole, ${\ket j}_L$ and ${\ket j}_R$ are states in the left and right CFT. The core of ER=EPR is the conjectured equivalence between entangled systems and wormhole geometries, i.e. that between any two entangled quantum systems there is a wormhole, possibly of Planckian size.\footnote{Susskind distinguishes between a modest and an ambitious version of the conjecture \cite{Susskind:2016jjb}. Modest ER=EPR is supposed to apply only to entangled black holes, while ambitious ER=EPR applies to any entangled systems. In this paper, we are concerned only with the modest version since we only talk about black holes.}
\begin{figure}[h!]
\centering
\includegraphics[scale=0.8]{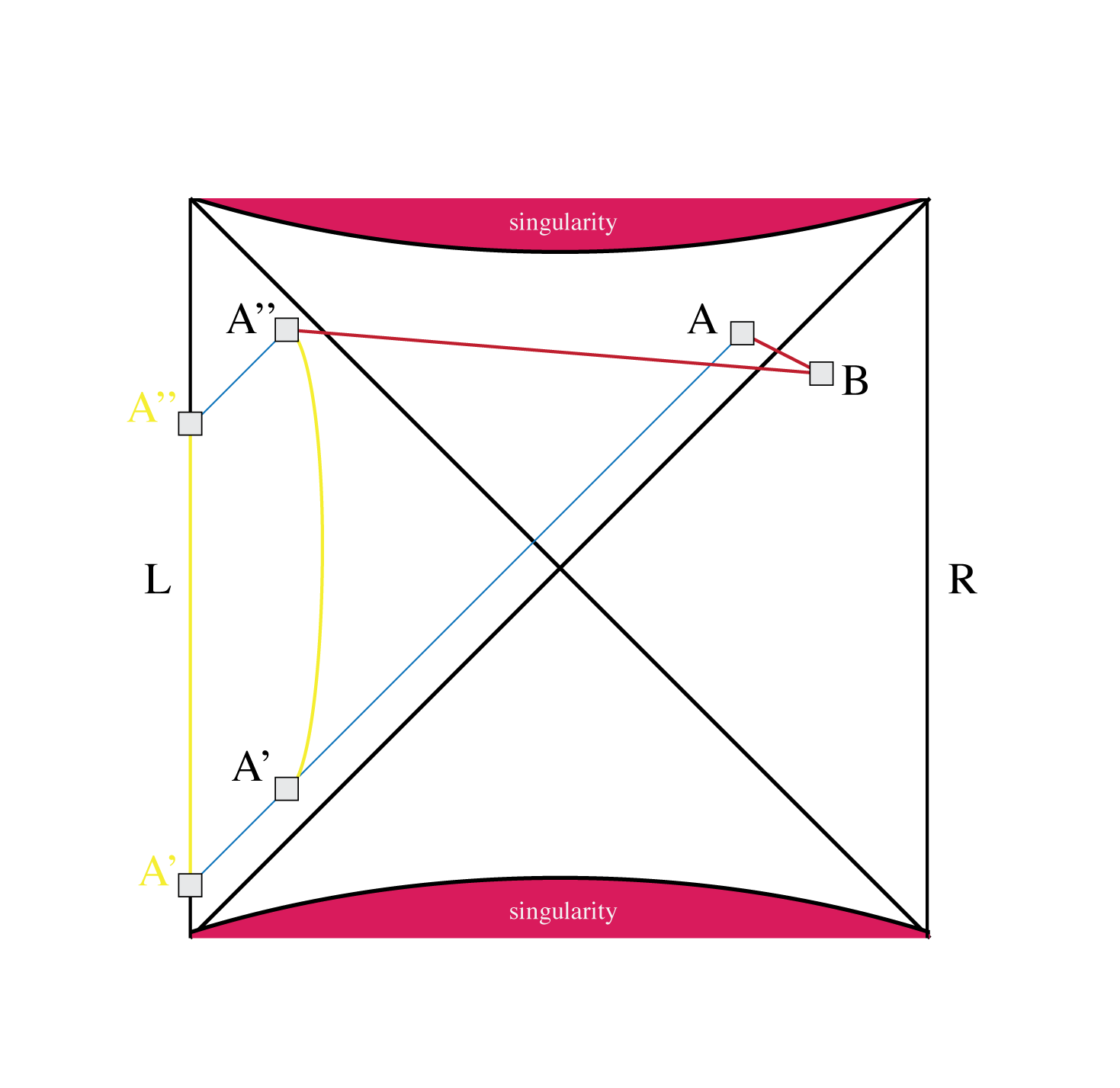} 
\caption{The Penrose diagram of an eternal AdS black hole (with $L$ the left exterior and $R$ the right exterior) with an AMPS situation and its ER=EPR resolution. Here red lines represent entanglement and blue lines semiclassical bulk evolution, while yellow lines represent boundary evolution and its bulk dual. Furthermore, $A$ is an interior mode, $A'$ is its CPT transform (with its holographic dual in yellow), and $A''$ is a mode on the left horizon (with its holographic dual in yellow) obtained by evolving $A'$. The AMPS-like situation comes from the maximal entanglement between $A''$ and $B$, and $A$ and $B$. The resolution is given by the existence of a unitary connection between $A''$ and $A$.}
\label{fig:ER=EPR}
\end{figure}
The eternal AdS black hole has a dual interpretation: we can either understand it as a system made of two entangled black holes or as two black holes connected by a wormhole. This dual way of looking at eternal AdS black holes is at the heart of the ER=EPR proposal. Indeed, the eternal AdS black hole is a particularly special case of the ER=EPR conjecture, in the sense that we do not need to modify the classical geometry of the black hole to get ER=EPR. As it were, the wormhole is already there in the eternal AdS black hole. In particular, no quantum wormhole is needed to verify the conjecture in this case, only classical geometry, making the subsequent discussion much easier. The basic lessons that we learn in this case, however, carry over also to the more realistic cases that we treat later, where no such convenient semiclassical picture is available.\\
The semiclassical features of the eternal AdS black hole are at the heart of how ER=EPR solves the AMPS paradox. To see why let us start by constructing a firewall like situation in the context of the eternal AdS black hole ({Figure} \ref{fig:ER=EPR}). We start at time $t$ by taking a pair of entangled qubits $A$ and $B$, on the right side of the two-sided eternal black hole, with $A$ behind the black hole horizon and $B$ outside the horizon. Let us now apply a CPT transformation\footnote{CPT here stands for \textit{charge}, \textit{parity} and \textit{time reversal}. CPT transformations are among the fundamental symmetries of quantum systems, such as the one we are considering here.} to $A$. {Since this transformation is a symmetry, sends $t$ into $-t$, and exchanges left and right, we get a qubit $A'$, equivalent to $A$, at time $-t$ in the left exterior region of the black hole.} Furthermore, it is clear from the Penrose diagram, that we can evolve $A'$ into $A$ with the bulk equations of motion. Let us write this as $A' \rightarrow A$. In particular, $A'$ is entangled with $B$ since we just applied a CPT transformation to $A$, and $A$ is entangled with $B$. Since $A$ is a qubit, it has three components {$A_i$ where $i=1,2,3$, which are Pauli matrices}. We thus have that:
\begin{equation}\label{eq5}
[A_i, A_j] = i\epsilon_{ijk}A_k \neq 0
\end{equation}
Since, however, by forwards time evolution with the bulk equations of motion, we have that $A'$ evolves in $A$, we can also write:
\begin{equation}\label{eq6}
[A'_i, A_j] \neq 0
\end{equation}
\noindent
To arrive at this result, we have relied on the fact that we can view the eternal AdS black hole as a wormhole connecting two horizons. A crucial step was the forwards time evolution of $A'$ in $A$, relying on $A'$ passing through the wormhole, as evident in {Figure} \ref{fig:ER=EPR}. However, we also know that we can regard the eternal AdS black hole as two entangled black holes, which means that we can evolve qubits in the left exterior (dual to the left CFT) independently from the right exterior (dual to the right CFT) since these are just two disconnected spacetimes. In particular, if we evolve $A'$ forwards in time up to time $t$, with the left CFT Hamiltonian, we obtain the qubit $A''$, which is naturally understood as the holographic dual to a qubit living on the left horizon.\footnote{To be precise, the left stretched horizon, though this is not relevant to the present discussion.} Since $A''$ is just the product of evolving forwards in time $A'$, it carries the same information (in the sense that they are related by a unitary operator, the left CFT Hamiltonian). In particular, it is entangled with $B$, giving us a firewall like situation. At $t$, we have a qubit $B$ entangled with a qubit in the interior ($A$) and a distant, far away qubit ($A''$). Here $B$ is equivalent to $L$ in our formulation of AMPS while $A$ is equivalent to $B$, and $A''$ to $E$ (remember {Figure} \ref{fig:AMPS}). It would thus seem that we have here too a violation of the monogamy of entanglement. However, we can immediately see the resolution of this apparent paradox. Since $A''$ follows from applying the left CFT Hamiltonian to $A'$, and $A' \rightarrow A$, we can also write $A'' \rightarrow A$.\footnote{Intuitively, we can understand this as first evolving backwards in time from $A''$ to $A'$ with the left CFT Hamiltonian, and then forwards in time from $A'$ to $A$ with the bulk equations of motion.} But then we can substitute $A''$ for $A'$ in \eqref{eq6}, giving us:
\beq\label{eq7}
 [A''_i, A_j] \neq 0
\eeq
\eqref{eq7} resolves the apparent violation of monogamy that we have engineered since it tells us that $A$ and $A''$ are not independent qubits (since they do not commute). As such, it is not the case that $B$ is maximally entangled with two different systems, leading to a violation of the monogamy of entanglement. Since $A$ and $A''$ are not distinct systems, $B$ is entangled with only one system, represented by both $A$ and $A''$, in a fundamentally non-local manner. The mistake in the reasoning which led us to an apparent violation of monogamy was the (background) assumption of \textbf{(SD)}, which told us that two spacelike separated systems must commute, and thus be distinct. Since $A$ and $A''$ are spacelike separated, it was only natural to assume that they were two different, distinct systems. What our analysis shows is that, in the context of the eternal AdS black hole where ER=EPR is already a feature of the semiclassical geometry, the assumption of \textbf{(SD)} falls apart, which is the essence of (iii)'s resolution of the AMPS paradox. Even spacelike separated, distant objects can still nonetheless depend on each other, and thus not be distinct objects after all.\footnote{{It would be an interesting project to better understand how the notions of distinctness and dependence that we use here can be related to the notions of distinctness and dependence analyzed in \cite{Schaffer2016-SCHGIT-2} in the framework of metaphysical grounding. Such a project, however, goes beyond the scope of the present paper.}} \\
This analysis is best understood in the language of causal structures \textbf{(CS)}. In $\S\ref{3.4}$, we have seen that the monogamy paradox can be recast as the mismatch between the causal structure of spacetime and the causal structure of entanglement. Since $R_{ME}(A,B)$ and $R_{ME}(A'',B)$ while $\neg R_{LC}(A'',A)$, which, by \textbf{(SD)}, implies that $A''$ and $A$ are distinct systems, we come to the conclusion that $B$ is maximally entangled with two distinct systems, violating monogamy. The way ER=EPR allows us to resolve the paradox, in the context of the eternal AdS black hole, is by defining a more general causal structure, characterized by the relation $R_{WH}$ of \textit{being non trivially connected}, which obtains if and only if two entities $A$ and $B$ can non trivially influence each other. A non-trivial influence is manifested by the presence of counterfactually robust correlations between $A$ and $B$. We call this the \textit{generalized causal structure}. \\
Observe that it is always the case that we can embed the causal structure of spacetime and entanglement in the generalized causal structure, by observing that both $R_{ME}$ and $R_{LC}$ are supposed to produce robust counterfactual correlations. However, and here lies causal structures' usefulness in understanding the ER=EPR resolution of AMPS, in certain situations such an embedding is not an isomorphism, i.e. there are systems connected by $R_{WH}$ which are connected neither by $R_{ME}$ nor $R_{LC}$. Indeed, in the eternal black hole in AdS, the wormhole connects the left and right exterior, making it possible to have a non-trivial connection between $A$ and $A''$, despite their not being entangled and being spacelike related. Indeed, it is this non-trivial connection, captured by $R_{WH}$, which leads to the violation of \textbf{(SD)}, as per (iii). \\
The eternal AdS black hole serves as a simple motivating case to understand how violations of \textbf{(SD)} naturally emerge in the study of black holes. We can now move on, in the next section, to the study of evaporating black holes formed from gravitational collapse. Let us, however, remark once more the most critical intuition underlying the ER=EPR conjecture: it is the idea that the overall structure of spacetime, as it emerges from QG, is much more complicated than its \textit{naive} semiclassical description would lead us to believe. There are many more connections that are not accounted for by merely thinking in terms of causal curves, and these connections are central to a proper understanding of black holes. Furthermore, while in some lucky cases (such as the eternal AdS black hole) we can understand these connections in geometrical terms as wormhole geometries, in other cases this is not possible. Instead, we have to resort to the more general idea of there being connections absent in the semiclassical description, imaginatively called \textit{planckian} or \textit{quantum} wormholes in \cite{Maldacena:2013xja}. These non-local (from the perspective of semiclassical spacetime) connections are the heart of the ER=EPR conjecture and lie at the core of the constructions described here.

\subsection{Evaporating Black Holes from Gravitational Collapse}\label{4.3}
\begin{figure}[h!]
\centering
\includegraphics[scale=0.5]{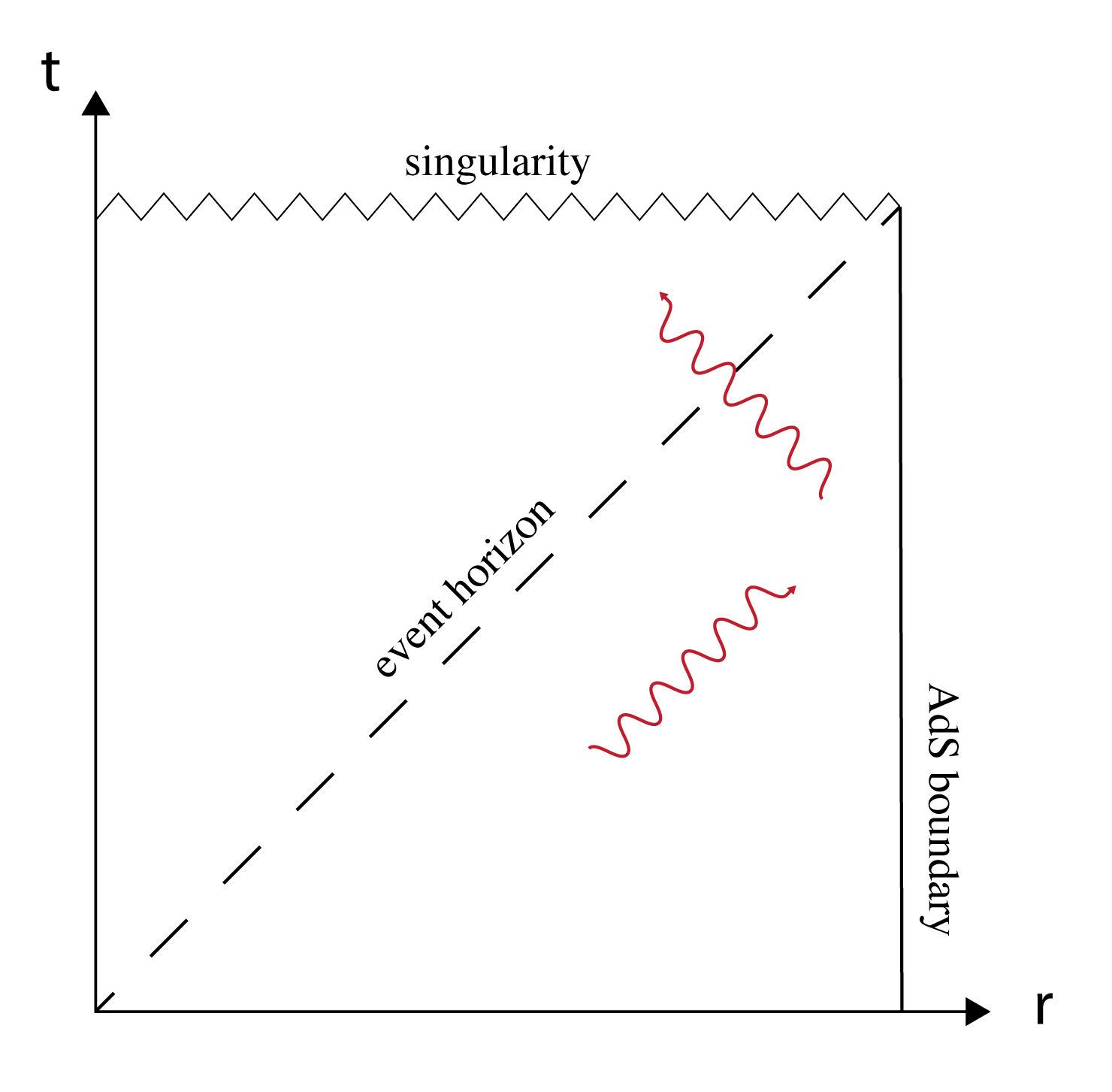} 
\caption{The Penrose diagram for a black hole with reflecting boundary conditions. Observe that the radiation (wiggly red lines) bounces on the boundary and comes back to the black hole.}
\label{fig:reflecting}
\end{figure}

\begin{figure}[h!]
\centering
\includegraphics[scale=0.60]{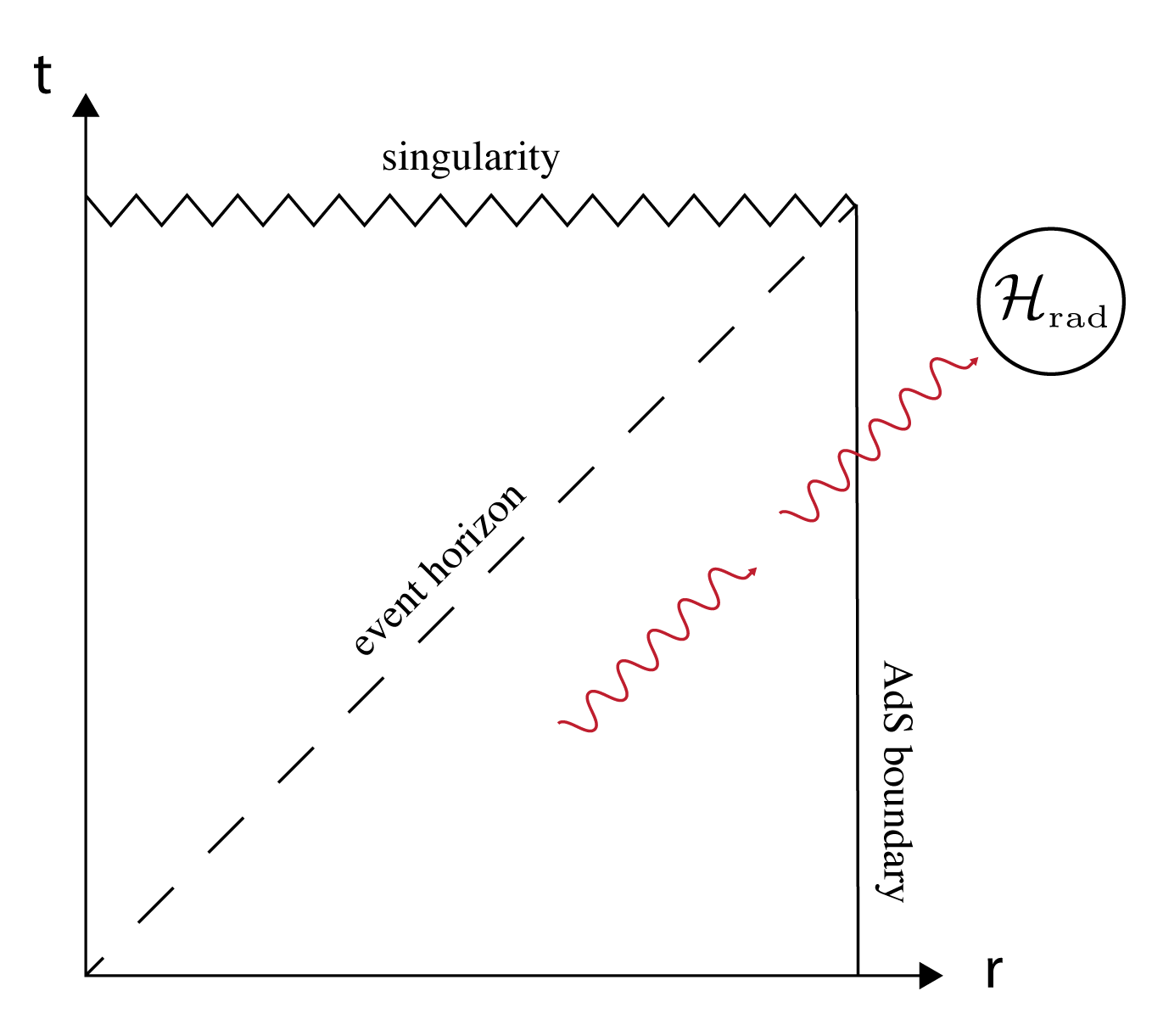} 
\caption{The Penrose diagram for a black hole with absorbing boundary conditions. Observe that some of the radiation (wiggly red lines) escapes from the boundary and is holographically encoded in $\mathcal H_{rad}$.}
\label{fig:absorbing}
\end{figure}

In this section, we study the firewall paradox in a situation where the black hole is formed from gravitational collapse and can evaporate, a problem studied extensively in \cite{penington2019entanglement}.\\
As we already said, the black hole in AdS spacetime considered in \S\ref{4.1} is not evaporating, since it has reflecting boundary conditions, and cannot give a real example of the firewall paradox. Indeed, the conformal boundary of AdS reflects the Hawking quanta into the black hole, reaching thermal equilibrium since the number of emitted and reflected quanta compensate (see {Figure} \ref{fig:reflecting}).
\\
A strategy one can consider to build an evaporating black hole is the following: take a black hole formed from collapse and place it into a spacetime whose boundary is not completely reflecting, i.e. in a spacetime that permits some Hawking quanta to escape outside the AdS boundary. In this case, the emitted Hawking radiation will be larger than the Hawking radiation coming back into the black hole, since some of the radiation has escaped outside the boundary. The black hole, then, slowly evaporates. The more radiation we permit to escape from AdS, the faster the black hole evaporates. This procedure can be made precise within the context of AdS/CFT by coupling the boundary CFT to an auxiliary reservoir $\mathcal H_{rad}$, using absorbing boundary conditions. Furthermore, we assume that $\mathcal H_{rad}$ is a large holographic system, which allows the (holographic) encoding of the Hawking radiation into $\mathcal H_{rad}$ (see {Figure} \ref{fig:absorbing}).
How can we incorporate $H_{rad}$, and thus the escaping Hawking radiation, into the analysis of $\S\ref{4.1}$? Take for instance a quantum of late Hawking radiation $L$, which, after the Page time, is entangled with some interior modes $B$ and with the early Hawking radiation $E$, thus having a monogamy problem. As in the two-sided black hole case of \S\ref{4.1}, where the interior mode $A$ was encoded in the left CFT, the interior mode $B$ is encoded in the boundary theory. In particular, for one-sided evaporating black holes, the mode $B$ is encoded in $\mathcal H_{rad}$.\footnote{This statement is the core of the analysis of \cite{penington2019entanglement}. Indeed \cite{penington2019entanglement} proves this claim using entanglement wedge reconstruction.} Before the Page time $t_p$, $\mathcal H_{rad}$ encodes only the early Hawking radiation that escapes from the boundary of AdS. After the Page time, $t_p$, $\mathcal H_{rad}$ also encodes $B$, despite it being in the interior and thus unable to reach the boundary. This fact signals the breakdown of the semiclassical picture since we have two spacelike separated systems $B$ and $E$, which live in the same CFT $\mathcal H_{rad}$. Thus, there is no problem for the late Hawking radiation $L$ to be entangled with both the interior mode $B$ and $\mathcal H_{rad}$ (which encodes the escaped early radiation $E$), since, as in the case of the eternal AdS black hole, the first statement implies the second one. Since the interior mode $B$ is encoded in $\mathcal H_{rad}$, entanglement with $B$ implies entanglement with a mode in $\mathcal H_{rad}$. Equivalently, \textbf{(SD)} is violated since we have two spacelike separated systems, one inside and one outside the black hole, $B$ and $\mathcal H_{rad}$ (which encodes $E$) which are nonetheless not distinct systems. Their not being distinct is a consequence of the fact that $B$ is encoded in $\mathcal H_{rad}$, which equivalently means that $B$ is a part of $\mathcal H_{rad}$. Thus, the two systems cannot be distinct.\\
From the perspective of causal structures, we can analyze this situation in the same way in which we have studied the eternal AdS black hole of \S\ref{4.1}. The paradox is that it seems to be the case that $R_{ME}(L,B)$ and $R_{ME}(L,E)$, while $\neg R_{LC}(E,B)$, which by \textbf{(SD)} implies that $B$ and $E$ are two distinct systems, violating monogamy. However, as we have seen in this section, although $E$ and $B$ are spacelike separated, they are not distinct. Indeed, after $t_p$, $B$ is holographically encoded in $\mathcal H_{rad}$ and therefore connected with $E$. This new connection can be encoded via $R_{WH}$ and corresponds to a connection which is captured neither by $R_{LC}$ nor by $R_{ME}$. Again, the generalized causal structure (defined via $R_{WH}$) captures the structure of the black hole that goes beyond the semiclassical approximation, which relies only on $R_{LC}$ and $R_{ME}$. In particular, $R_{WH}$ encodes those connections which show that the two systems $B$ and $E$ which, from the perspective of the causal structure of spacetime and entanglement, are distinct and separated, are interdependent and connected, thus violating \textbf{(SD)}. \\
Furthermore, this is again the same intuition of ER=EPR, that the semiclassical picture of spacetime crucially fails in the context of black holes in taking into account non-local connections that are neither causal curves nor entanglement relations. One of the main advantages of the causal structure approach is that we can naturally show the underlying strategy behind the different proposals for resolving the AMPS paradox that we have studied thus far.

%%%%%%%%%%%%%%%%%%%%%%%%%%%%%%%%%%%%%%%%%%%%%%%%%%%%%%%%%%%%%%%%%%%%%%%%%%%%%%%%%%%%%%%%%%%%%%%%%%%%%%%%%%%%%%%%%%%%%%%%%%%%%%%%%%%%%%%%%%%%%%%%%%%%%%%%%%%%%%%%%%%%%%%%%%%%%%%%%%%%%%%%%%%%%%%%%%%%%%%%%%%%%%%%%%%%%%%%%%%%%%%%%%%%%%%%%%%%%%%%%%%%%%%%%%%%%%%%%%%%%%%%%%%%%%%%%%%%%%%%%%%%%%%%%%%%%%%%%%%%%%%%%%%%%%%%%%%%%%%

\section{Conclusions}\label{5}
{In this paper, we have seen how the AMPS paradox appears to threaten the consistency of black hole physics and how dropping the implicit assumption of \textbf{spacetime distinctness} allows us to overcome it. The core of our work is the notion of causal structure. In particular, this notion helped us in highlighting the role of \textbf{(SD)} and in making explicit the link between various strategies for constructing the interior of the black hole, as explained in \S\ref{4.1} and \S\ref{4.3}. Furthermore, causal structures make clear the sense in which ER=EPR-like connections imply a violation of semiclassical locality.\\
Interestingly enough \textbf{(SD)} being violated seems reminiscent of the basic idea of black hole complementarity \cite{Susskind_1993}, i.e. the idea that the interior and the exterior are two non-commuting descriptions of the same physics. It would then seem that the models that we have analyzed provide a precise incarnation of black hole complementarity by instantiating the violation of \textbf{(SD)}. However, understanding the connection between \textbf{(SD)} and black hole complementarity, and how the ideas of this paper can help black hole complementarity out of the disrepute in which many philosophers of physics hold it is beyond the scope of this paper and will be left for future works.\\
Furthermore, it is far from clear that the non-locality involved in the constructions we have studied is not problematic. A task of extreme importance is understanding whether this non-locality leaks in the low energy regime \cite{Marolf:2015dia}. Moreover, it is also essential to understand what it means for locality to be emergent. \\
Most importantly, we have limited ourselves in this work to the AMPS paradox.} However, AMPS is not the only paradox involved in constructing the interior of the black hole. Most notably, in the context of AdS/CFT, we have not discussed the AMPSS paradox \cite{Almheiri:2013hfa}\footnote{{Here AMPSS refers to a second black hole paradox, distinct from the AMPS paradox introduced in \cite{Almheiri_2013}. We will discuss the AMPSS paradox in a follow-up paper.}} and related problems in the field of holographic interior reconstruction \cite{Harlow:2018fse}. We have also not sufficiently compared dropping \textbf{(SD)} with the firewall solution of the paradox. Both these topics will be addressed in two follow up papers.

\section*{Acknowledgements}
For helpful comment on an a earlier draft we would like to thank Alberto Corti, Vincenzo Fano, Christian W\"uthrich, and the audience of the workshop \enquote{A New Perspective on the Paradoxes of Modern Physics}.

%%%%%%%%%%%%%%%%%%%%%%%%%%%%%%%%%%%%%%%%%%%%%%%%%%%%%%%%%%%%%%%%%%%%%%%%%%%%%%%%%%%%%%%%%%%%%%%%%%%%%%%%%%%%%%%%%%%%%%%%%%%%%%%%%%%%%%%%%%%%%%%%%%%%%%%%%%%%%%%%%%%%%%%%%%%%%%%%%%%%%%%%%%%%%%%%%%%%%%%%%%%%%%%%%%%%%%%%%%%%%%%%%%%%%%%%%%%%%%%%%%%%%%%%%%%%%%%%%%%%%%%%%%%%%%%%%%%%%%%%%%%%%%%%%%%%%%%%%%%%%%%%%%%%%%%%%%%%%%%

\appendix
\bibliographystyle{chicago}

\bibliography{Bibliography}

\begin{thebibliography}{}

\bibitem[\protect\citeauthoryear{Almheiri}{Almheiri}{2018}]{almheiri2018holographic}
Almheiri, A. (2018).
\newblock Holographic quantum error correction and the projected black hole
  interior.

\bibitem[\protect\citeauthoryear{Almheiri, Engelhardt, Marolf, and
  Maxfield}{Almheiri et~al.}{2019}]{Almheiri:2019psf}
Almheiri, A., N.~Engelhardt, D.~Marolf, and H.~Maxfield (2019).
\newblock {The entropy of bulk quantum fields and the entanglement wedge of an
  evaporating black hole}.
\newblock {\em JHEP\/}~{\em 12}, 063.

\bibitem[\protect\citeauthoryear{Almheiri, Hartman, Maldacena, Shaghoulian, and
  Tajdini}{Almheiri et~al.}{2020}]{Almheiri:2019qdq}
Almheiri, A., T.~Hartman, J.~Maldacena, E.~Shaghoulian, and A.~Tajdini (2020).
\newblock {Replica Wormholes and the Entropy of Hawking Radiation}.
\newblock {\em JHEP\/}~{\em 05}, 013.

\bibitem[\protect\citeauthoryear{Almheiri, Mahajan, Maldacena, and
  Zhao}{Almheiri et~al.}{2020}]{Almheiri:2019hni}
Almheiri, A., R.~Mahajan, J.~Maldacena, and Y.~Zhao (2020).
\newblock {The Page curve of Hawking radiation from semiclassical geometry}.
\newblock {\em JHEP\/}~{\em 03}, 149.

\bibitem[\protect\citeauthoryear{Almheiri, Marolf, Polchinski, Stanford, and
  Sully}{Almheiri et~al.}{2013}]{Almheiri:2013hfa}
Almheiri, A., D.~Marolf, J.~Polchinski, D.~Stanford, and J.~Sully (2013).
\newblock {An Apologia for Firewalls}.
\newblock {\em JHEP\/}~{\em 09}, 018.

\bibitem[\protect\citeauthoryear{Almheiri, Marolf, Polchinski, and
  Sully}{Almheiri et~al.}{2013}]{Almheiri_2013}
Almheiri, A., D.~Marolf, J.~Polchinski, and J.~Sully (2013, Feb).
\newblock Black holes: complementarity or firewalls?
\newblock {\em Journal of High Energy Physics\/}~{\em 2013\/}(2), 062.

\bibitem[\protect\citeauthoryear{Ammon and Erdmenger}{Ammon and
  Erdmenger}{2015}]{10.5555/2834415}
Ammon, M. and J.~Erdmenger (2015).
\newblock {\em Gauge/Gravity Duality: Foundations and Applications\/} (1st
  ed.).
\newblock USA: Cambridge University Press.

\bibitem[\protect\citeauthoryear{Belot, Earman, and Ruetsche}{Belot
  et~al.}{1999}]{10.2307/40072220}
Belot, G., J.~Earman, and L.~Ruetsche (1999).
\newblock The hawking information loss paradox: The anatomy of a controversy.
\newblock {\em The British Journal for the Philosophy of Science\/}~{\em
  50\/}(2), 189--229.

\bibitem[\protect\citeauthoryear{Curiel}{Curiel}{2019}]{Curiel:2018cbt}
Curiel, E. (2019).
\newblock {The many definitions of a black hole}.
\newblock {\em Nature Astron.\/}~{\em 3\/}(1), 27--34.

\bibitem[\protect\citeauthoryear{De~Haro, Mayerson, and Butterfield}{De~Haro
  et~al.}{2016}]{deHaro:2015pia}
De~Haro, S., D.~R. Mayerson, and J.~N. Butterfield (2016).
\newblock {Conceptual Aspects of Gauge/Gravity Duality}.
\newblock {\em Found. Phys.\/}~{\em 46\/}(11), 1381--1425.

\bibitem[\protect\citeauthoryear{Earman and Valente}{Earman and
  Valente}{2014}]{Earman2014-EARRCI}
Earman, J. and G.~Valente (2014).
\newblock Relativistic causality in algebraic quantum field theory.
\newblock {\em International Studies in the Philosophy of Science\/}~{\em
  28\/}(1), 1--48.

\bibitem[\protect\citeauthoryear{Einstein, Podolsky, and Rosen}{Einstein
  et~al.}{1935}]{PhysRev.47.777}
Einstein, A., B.~Podolsky, and N.~Rosen (1935, May).
\newblock Can quantum-mechanical description of physical reality be considered
  complete?
\newblock {\em Phys. Rev.\/}~{\em 47}, 777--780.

\bibitem[\protect\citeauthoryear{Einstein and Rosen}{Einstein and
  Rosen}{1935}]{PhysRev.48.73}
Einstein, A. and N.~Rosen (1935, Jul).
\newblock The particle problem in the general theory of relativity.
\newblock {\em Phys. Rev.\/}~{\em 48}, 73--77.

\bibitem[\protect\citeauthoryear{Giddings}{Giddings}{2013a}]{Giddings:2013kcj}
Giddings, S.~B. (2013a).
\newblock {Nonviolent information transfer from black holes: A field theory
  parametrization}.
\newblock {\em Phys. Rev. D\/}~{\em 88\/}(2), 024018.

\bibitem[\protect\citeauthoryear{Giddings}{Giddings}{2013b}]{Giddings:2012gc}
Giddings, S.~B. (2013b).
\newblock {Nonviolent nonlocality}.
\newblock {\em Phys. Rev. D\/}~{\em 88}, 064023.

\bibitem[\protect\citeauthoryear{Giddings}{Giddings}{2013c}]{Giddings:2013vda}
Giddings, S.~B. (2013c).
\newblock {Statistical physics of black holes as quantum-mechanical systems}.
\newblock {\em Phys. Rev. D\/}~{\em 88}, 104013.

\bibitem[\protect\citeauthoryear{Giddings}{Giddings}{2014}]{Giddings:2014ova}
Giddings, S.~B. (2014).
\newblock {Possible observational windows for quantum effects from black
  holes}.
\newblock {\em Phys. Rev. D\/}~{\em 90\/}(12), 124033.

\bibitem[\protect\citeauthoryear{Haag}{Haag}{2012}]{haag2012local}
Haag, R. (2012).
\newblock {\em Local quantum physics: Fields, particles, algebras}.
\newblock Springer Science \& Business Media.

\bibitem[\protect\citeauthoryear{Halvorson}{Halvorson}{2007}]{halvorson2007algebraic}
Halvorson, H. (2007).
\newblock Algebraic quantum field theory.
\newblock In J.~Earman and J.~Butterfield (Eds.), {\em Handbook of Philosophy
  of Physics}, pp.\  731--922. Elsevier.

\bibitem[\protect\citeauthoryear{Harlow}{Harlow}{2016}]{Harlow:2014yka}
Harlow, D. (2016).
\newblock {Jerusalem Lectures on Black Holes and Quantum Information}.
\newblock {\em Rev. Mod. Phys.\/}~{\em 88}, 015002.

\bibitem[\protect\citeauthoryear{Harlow}{Harlow}{2018}]{Harlow:2018fse}
Harlow, D. (2018).
\newblock {TASI Lectures on the Emergence of Bulk Physics in AdS/CFT}.
\newblock {\em PoS\/}~{\em TASI2017}, 002.

\bibitem[\protect\citeauthoryear{Harlow and Hayden}{Harlow and
  Hayden}{2013}]{Harlow_2013}
Harlow, D. and P.~Hayden (2013, Jun).
\newblock Quantum computation vs. firewalls.
\newblock {\em Journal of High Energy Physics\/}~{\em 2013\/}(6), 085.

\bibitem[\protect\citeauthoryear{Hawking}{Hawking}{1976}]{PhysRevD.14.2460}
Hawking, S.~W. (1976, Nov).
\newblock Breakdown of predictability in gravitational collapse.
\newblock {\em Phys. Rev. D\/}~{\em 14}, 2460--2473.

\bibitem[\protect\citeauthoryear{Hayden and Penington}{Hayden and
  Penington}{2019}]{Hayden:2018khn}
Hayden, P. and G.~Penington (2019).
\newblock {Learning the Alpha-bits of Black Holes}.
\newblock {\em JHEP\/}~{\em 12}, 007.

\bibitem[\protect\citeauthoryear{Jarrett}{Jarrett}{1984}]{jarrett1984physical}
Jarrett, J.~P. (1984).
\newblock On the physical significance of the locality conditions in the bell
  arguments.
\newblock {\em No{\^u}s\/}~{\em 18\/}(4), 569--589.

\bibitem[\protect\citeauthoryear{Maldacena and Susskind}{Maldacena and
  Susskind}{2013}]{Maldacena:2013xja}
Maldacena, J. and L.~Susskind (2013).
\newblock {Cool horizons for entangled black holes}.
\newblock {\em Fortsch. Phys.\/}~{\em 61}, 781--811.

\bibitem[\protect\citeauthoryear{Maldacena}{Maldacena}{1999}]{Maldacena:1997re}
Maldacena, J.~M. (1999).
\newblock {The Large N limit of superconformal field theories and
  supergravity}.
\newblock {\em Int. J. Theor. Phys.\/}~{\em 38}, 1113--1133.

\bibitem[\protect\citeauthoryear{Marolf and Polchinski}{Marolf and
  Polchinski}{2016}]{Marolf:2015dia}
Marolf, D. and J.~Polchinski (2016).
\newblock {Violations of the Born rule in cool state-dependent horizons}.
\newblock {\em JHEP\/}~{\em 01}, 008.

\bibitem[\protect\citeauthoryear{Maudlin}{Maudlin}{1994}]{Maudlin2002-MAUQNA}
Maudlin, T. (1994).
\newblock {\em Quantum Non-Locality and Relativity: Metaphysical Intimations of
  Modern Physics}.
\newblock Blackwell.

\bibitem[\protect\citeauthoryear{Maudlin}{Maudlin}{2017}]{Maudlin:2017lye}
Maudlin, T. (2017).
\newblock {(Information) Paradox Lost}.
\newblock last accessed 16-09-2020.

\bibitem[\protect\citeauthoryear{Mussardo}{Mussardo}{2010}]{mussardo2010statistical}
Mussardo, G. (2010).
\newblock {\em Statistical field theory: an introduction to exactly solved
  models in statistical physics}.
\newblock Oxford University Press.

\bibitem[\protect\citeauthoryear{Page}{Page}{1993}]{Page_1993}
Page, D.~N. (1993, Dec).
\newblock Information in black hole radiation.
\newblock {\em Physical Review Letters\/}~{\em 71\/}(23), 3743--3746.

\bibitem[\protect\citeauthoryear{Papadodimas and Raju}{Papadodimas and
  Raju}{2013}]{Papadodimas:2012aq}
Papadodimas, K. and S.~Raju (2013).
\newblock {An Infalling Observer in AdS/CFT}.
\newblock {\em JHEP\/}~{\em 10}, 212.

\bibitem[\protect\citeauthoryear{Papadodimas and Raju}{Papadodimas and
  Raju}{2016}]{Papadodimas:2015jra}
Papadodimas, K. and S.~Raju (2016).
\newblock {Remarks on the necessity and implications of state-dependence in the
  black hole interior}.
\newblock {\em Phys. Rev. D\/}~{\em 93\/}(8), 084049.

\bibitem[\protect\citeauthoryear{Penington}{Penington}{2019}]{penington2019entanglement}
Penington, G. (2019).
\newblock Entanglement wedge reconstruction and the information paradox.

\bibitem[\protect\citeauthoryear{Rovelli}{Rovelli}{2019}]{Rovelli_2019}
Rovelli, C. (2019, Aug).
\newblock The subtle unphysical hypothesis of the firewall theorem.
\newblock {\em Entropy\/}~{\em 21\/}(9), 839.

\bibitem[\protect\citeauthoryear{Schaffer}{Schaffer}{2016}]{Schaffer2016-SCHGIT-2}
Schaffer, J. (2016).
\newblock Grounding in the image of causation.
\newblock {\em Philosophical Studies\/}~{\em 173\/}(1), 49--100.

\bibitem[\protect\citeauthoryear{Susskind}{Susskind}{2008}]{susskind2008black}
Susskind, L. (2008).
\newblock {\em The Black Hole War: My Battle with Stephen Hawking to Make the
  World Safe for Quantum Mechanics}.
\newblock Little, Brown.

\bibitem[\protect\citeauthoryear{Susskind}{Susskind}{2016}]{Susskind:2016jjb}
Susskind, L. (2016).
\newblock {Copenhagen vs Everett, Teleportation, and ER=EPR}.
\newblock {\em Fortsch. Phys.\/}~{\em 64\/}(6-7), 551--564.

\bibitem[\protect\citeauthoryear{Susskind, Thorlacius, and Uglum}{Susskind
  et~al.}{1993}]{Susskind_1993}
Susskind, L., L.~Thorlacius, and J.~Uglum (1993, Oct).
\newblock The stretched horizon and black hole complementarity.
\newblock {\em Physical Review D\/}~{\em 48\/}(8), 3743--3761.

\bibitem[\protect\citeauthoryear{Unruh and Wald}{Unruh and
  Wald}{1995}]{Unruh:1995gn}
Unruh, W.~G. and R.~M. Wald (1995).
\newblock {On evolution laws taking pure states to mixed states in quantum
  field theory}.
\newblock {\em Phys. Rev. D\/}~{\em 52}, 2176--2182.

\bibitem[\protect\citeauthoryear{Unruh and Wald}{Unruh and
  Wald}{2017}]{Unruh:2017uaw}
Unruh, W.~G. and R.~M. Wald (2017).
\newblock {Information Loss}.
\newblock {\em Rept. Prog. Phys.\/}~{\em 80\/}(9), 092002.

\bibitem[\protect\citeauthoryear{Wallace}{Wallace}{2020}]{Wallace:2017wzs}
Wallace, D. (2020, 4).
\newblock {Why Black Hole Information Loss is Paradoxical}.
\newblock In N.~Huggett, K.~Matsubara, and C.~W\"uthrich (Eds.), {\em {Beyond
  Spacetime}}, pp.\  209--236. Cambridge University Press.

\end{thebibliography}

\end{document}